# Studying Cascading Overload Failures under High Penetration of Wind Generation

Mir Hadi Athari
Department of Electrical and Computer Engineering
Virginia Commonwealth University
Richmond, VA, USA
Email: atharih@vcu.edu

Zhifang Wang
Department of Electrical and Computer Engineering
Virginia Commonwealth University
Richmond, VA, USA
Email: zfwang@vcu.edu

*Abstract*—While power systems are reliable infrastructures, their complex interconnectivities allow for propagation of disturbances through cascading failures which causes blackouts. Meanwhile the ever increasing penetration level of renewable generation into power grids introduces a massive amount of uncertainty to the grid that might have a severe impact on grid vulnerability to overload cascading failures. There are numerous studies in the literature that focus on modeling cascading failures with different approaches. However, there is a need for studies that simulate cascading failure considering the uncertainty coming from high penetration of renewable generation. In this study, the impacts of wind generation in terms of its penetration and uncertainty levels on grid vulnerability to cascading overload failures are studied. The simulation results on IEEE 300 bus system show that uncertainty coming from wind energy have severe impact on grid vulnerability to cascading overload failures. Results also suggest that higher penetration levels of wind energy if not managed appropriately will add to this severity due to injection of higher uncertainties into the grid.

## I. INTRODUCTION

Electric power systems are drivers for modernity and sustainable growth in societies and they are considered as critical infrastructure requiring the highest levels of reliability. However, large-scale blackouts resulting from cascading failures impose a massive cost to societies annually. These cascading failures originate from strong interdependencies inside power grids and can even propagate into other critical infrastructures such as telecommunication, transportation, and water supply [1], [2]. Massive economic and social impacts of such events have motivated a great deal of research effort on studying the vulnerability of the power grids to cascading failures. A sequence of dependent failures of individual components that successively weakens the power system is defined as cascading failure [3]. This phenomenon can be result of different initial causes such as voltage and frequency instability, protection systems hidden failures, operator error, and transmission line overloads resulting from contingencies in the network. Each of these initial causes requires a different modelling technique and implementation of special approaches to evaluate grid vulnerability for certain events. Cascading failure due to line overloads has been under investigation by many researchers over past two decades. Probabilistic approach is among the mostly used technique to model the cascading behavior [4]–[9].

On the other hand, ever increasing penetration level of renewable energy sources into power grids will introduce a large amount of uncertainty coming from their stochastic nature. The dynamic performance and economic settings of many power systems are changing due to the increased penetrations of renewables [10]. One worrisome change is the increase in cascading failures involving wind farms [11].

With the sustained and rapid growth of renewable generation, their influence on power systems dynamic performance would be significant, therefore, studying cascading overload failures considering uncertainty coming from renewable energy and loads is of great importance. In [9] authors proposed a stochastic Markov model to capture the progression of cascading failures considering uncertainty coming from electrical loads. However, the unpredictable nature of renewable generation such as wind energy will inject a large portion of uncertainty into the grid having a big impact on cascading failure behavior. A new uncertainty modeling approach for renewable generation and smart grid loads proposed in [12] which enables us to study the impacts of increased penetration level of renewable generation on grid vulnerabilities.

In this paper, we investigate the impacts of increased penetration of wind generations on cascading overload failure behavior in power grids. The main focus is on the uncertainty coming from these sources and as well as electrical loads. Simulations are performed on IEEE 300 bus system for different scenarios. In the proposed cascading failure model, the time space correlation of line flows uncertainty is taken into account by calculating the covariance matrix of flow uncertainty.

The rest of the paper is organized as follows. In section II the uncertainty modeling is briefly discussed. Cascading failure modeling procedure and techniques used are presented in section III. Section IV and V will present results and conclusion, respectively.

## II. UNCERTAINTY MODELING

Uncertainties coming from different sources such as renewable generation and loads present different characteristics in terms of magnitude and frequency of occurrence. In the model proposed in [12], the power injection from each component (i.e. output power for generators and demand power for loads) is modeled with two terms as shown in (1).

$$P(t) = \mu_P(t) + \epsilon_P(t) \quad (1)$$

where $\mu_P(t)$ is the average of power signal at each time instant or in other words it is what we expect to have for each component ahead of time and $\epsilon_P(t)$ which is a zero mean signal, represents the uncertainty which may come from forecast error or mismatch in output power for conventional units.

Using the power spectral density (PSD) of error signal, the occupied bandwidth for each uncertainty signal is calculated. It is found that wind generation injects the highest amount of uncertainty into the grid in terms of its bandwidth and magnitude and assumes all the bus voltages in the grid network closed to the rated voltage levels [12]. In the next section, the implementation of above model in cascading failure simulation will be explained.

## III. CASCADING FAILURE SIMULATION

### A. DC power flow and line overload modeling

Determining the steady-state operating conditions of the power grid requires solving the full power flow equations that provide information about the voltage magnitudes and phases and the active and reactive power flows through each transmission line. Unfortunately, solving repeatedly the full nonlinear power flow equations becomes computationally prohibitive due to numerous solutions during the evolution of CF. In addition, we are only interested in evaluating the impact of wind uncertainty on grid vulnerability to CF that doesn't necessarily require complete nonlinear network equations and the linear approximation for line active flows is sufficient for our tripping mechanism. For all these reasons we used DC power flow approximation [13] to recalculate the flow dispatch of the power grid at each time. Simulation of cascading failure is performed on IEEE 300 bus system replacing several conventional generators of the original system with wind generators according to different scenarios. The power flow equations for a power grid with $n$ nodes and $m$ links can be expressed as:

$$P(t) = B'(t)\theta(t) \quad (2)$$

$$F(t) = diag(y_l(t))A\theta(t) \quad (3)$$

where $P(t)$ represents the vector of injected real power, $\theta(t)$ the nodal voltage angles, and $F(t)$ the flows on the lines. The matrix $B'(t)$ is defined as

$$B'(t) = A^T diag(y_l(t))A \quad (4)$$

where $y_l(t) = 1/x_l$ is the line admittance; $diag(y_l(t))$ represents a diagonal matrix with entries of $\{y_l(t), l = 1,2,...,m\}$. $A \coloneqq (A_{l,k})_{m \times n}$ is the line-node incidence matrix, arbitrarily oriented and defined as: $A_{l,i} = 1$; $A_{l,j} = -1$, if the $l$th line is from node $i$ to node $j$ and $A_{l,k} = 0, k \neq i,j$.

Using the uncertainty modeling mentioned earlier, the injected power at each bus would be presented with two terms. Then using this representation we can form the mean and covariance matrices. The covariance matrix is calculated for each time step using the sliding window method as a $m \times m$ matrix:

$$C_F(t,\tau) = E\{\epsilon_F(t)\epsilon_F(t-\tau)\} = \begin{bmatrix} C_{11} & \cdots & C_{1m} \\ \vdots & \ddots & \vdots \\ C_{m1} & \cdots & C_{mm} \end{bmatrix} \quad (5)$$

$$C_{ij} = cov(\epsilon_{F_i}(t), \epsilon_{F_j}(t-\tau)) \quad (6)$$

$$cov(A,B) = E\{(A - \mu_A)(B - \mu_B)\} \quad (7)$$

where $\epsilon_{F_l}(t)$ is the uncertainty term in flow signal of line '$l$' at time $t$ and $cov$ is Covariance function defined in (7). In the simulation we utilize the unbiased estimates as $cov(A,B) = \frac{1}{N-1}\sum_{i=-N/2}^{N/2}(A_i - \mu_A)^*(B_i - \mu_B)$. Note that sliding window method for covariance calculation selects a predetermined number ($N$) of observations of line flow errors and then puts them in the observation matrix. By using eq. (8), each element of covariance matrix is calculated based on the observation matrix. The variance for flow process of each line can then be calculated by taking square root of each diagonal element in covariance matrix with $\sigma_{F_l}(t) = \sqrt{C_{F_{l,l}}(t,0)}$.

With a Gaussian assumption for the distribution of $F_l(t)$ in [8, 11], the overloading probability $\rho_l(t) = p\{|F_l(t)| > F_l^{max}\}$ can be calculated using Q-function as below:

$$\rho_l(t) \cong Q(a_l) \quad (8)$$

where $a_l = \frac{F_l^{max} - \mu_{F_l}(t)}{\sigma_{F_l}(t)}$ is the *normalized overload distance* of the $l$th line and Q-function as $Q(x) = \int_x^\infty e^{-t^2/2}/(\sqrt{2\pi})dt$.

Finally, using the normalized overload distance ($a_l$) and overloading probability ($\rho_l$) for each line we can calculate the mean overload time for flow process $F_l(t)$ as follows [9]:

$$\bar{\tau}_l^u = \frac{2\pi \rho_l e^{a_l^2/2}}{BW_l} \quad (9)$$

where $BW_l$ is the equivalent bandwidth of the flow process for the $l$th line and can be calculated using the spectral power density (SPD) of flow process [12].

### B. Tripping mechanism and relay model

The trip time of thermal overload relays is determined based on maximum allowable current flowing in the conductor without causing thermal instability. Generally, the overload relays for HV transmission lines have time-dependent tripping characteristic, which is determined using the well-known dynamic thermal balance between heat gains and losses in the conductor [14]. The maximum or hot spot temperature determines the time to trip for thermal relays and considering initial operation current and applying necessary changes the time to trip can be calculated using [15]:

$$t_{trp} = T_{th} \cdot \ln\left(\frac{F^2 - F_{op}^2}{F^2 - F_{max}^2}\right) \quad (10)$$

where $F$ is overload line flow (p.u.), $F_{op}$ is initial operating flow (p.u.), $F_{max}$ is the line flow threshold, and $T_{th}$ is thermal time constant which is related to conductor type and environmental parameters such as wind speed and ambient temperature [16]. In this study, it is assumed that all transmission lines use typical HAWK (477 kcmil) ACSR conductor with $T_{th} = 450$ sec.

In this study, for the tripping mechanism both relay time to trip and overloading probability are considered simultaneously to select the most probable line trip during the escalation phase of cascading failure. At every time step, first the time to trip for each overloaded line is calculated, then using normalized overload distance ($a_l$) and overloading probability ($\rho_l$) the mean overload time ($\bar{\tau}_l^u$) is determined. If relay time to trip is larger than the mean overload time, the trip timer is set to zero, otherwise the trip timer is set to the relay time to trip. This tripping mechanism enables us to model the stochastic process of cascading failure and identify the most probable path for its propagation.

## C. Island detection and power balance

Successive line tripping during escalation phase of cascading failure usually leads to formation of several islands in power network. The electrical frequency of the system is driven based on the power balance according to the well-known electro-mechanical equation [17]

$$\sum P_{gen} - \sum P_{load} = H. 2\pi \frac{df}{dt} \qquad (11)$$

where $\sum P_{gen}$ is the total produced power, $\sum P_{load}$ is the total consumed power, $H$ is the global inertia, and $f$ is the electrical frequency of the system. The frequency in power systems is considered a global parameter and cannot be influenced by a small section. However, when a part of network becomes islanded, the inertia and the load balance depend only on generators and loads inside the island where shedding actions may be necessary.

In the power balance algorithm for any island, the total load and total generation capacity are compared to each other. If the total demand exceeds the maximum available generation, some load shedding is necessary to maintain the power balance. Similarly, if total demand is smaller than the current generation, generation units should drop their output power.

An automatic island detection algorithm inspired by the approach proposed in [17] is used after each trip to identify newly formed islands. Clusters of generator(s) and load(s) that are not connected to the grid are called island(s). The algorithm consists of three steps; connectivity check, critical events identification, and island identification. Connectivity check determines how many islands are present in the power system, and their structure. Critical event detection identifies which breakers should create an island if opened. And the final step, island detection, identifies the buses belonging to each possible island and calculates their load balance. The actual dispatch of the network is not required for the algorithm since it only depends on gird topology (grid incidence matrix $A$) and generators location. Assuming equal 1 Ω resistance for all lines in the network and using the Kirchhoff current and voltage laws, the equations describing system behavior are calculated. To detect buses belonging to each island, generators are activated (assuming output current of 1 A) one at a time. After identifying all present separate islands in the grid, their power balance is maintained by shedding actions. For detailed island detection algorithm please refer to [17].

## IV. RESULTS & DISCUSSIONS

The simulation of cascading failure behavior considering wind generation uncertainty is performed on the IEEE 300 bus system by replacing some of its conventional generators with wind generators in two different scenarios. The IEEE 300 bus system is a synthesized network from New England power system and has a topology with 300 buses and 411 transmission lines. In the first scenario 11 conventional generators are replaced with wind generators at buses 80,88,125,128,156,199,222,256,258,262,295. The load and generation data are received from Electric Reliability Council of Texas (ERCOT). Uncertainty modeling of loads and generations are based on the model proposed in [12] and using Autoregressive Moving Average (ARMA) forecasting technique.

The initial operating equilibrium and conditions $(G(0), L(0), \theta(0), F(0))$ are taken or derived from the power flow solution. The equivalent bandwidth of flow process for each line under the initial uncertainty level is then calculated and stored to use later on stochastic tripping mechanism. The line capacities are set as $F^{max} = max\{\eta|F(0)|, 2.0(p.u.)\}$ with $\eta = 1.20$. Here we take $F(0)$ as the rational flow distribution under normal operating conditions and assume that the line capacity allows a 20% load increase [9]. The minimum of line capacity is set to be 2.0 p.u. so that the vibration in the lines which usually carry small flows will not cause frequent line trips.

First scenario is designed to study the impacts of forecasting relative error which comes into picture in the form of uncertainty from wind generation. For this scenario the uncertainty signal magnitude for wind generator is increased by factor $\gamma = \frac{\epsilon_w^{new}}{\epsilon_w^{int}}$ where $\epsilon_w^{new}$ is the new uncertainty term in wind power and $\epsilon_w^{int}$ is the initial uncertainty.

To see the impact of larger forecasting errors on grid vulnerability to overload cascading failures $\gamma$ is increased from 1 to 6 with 0.25 steps to find the uncertainty level in which the first cascading failure occurs. All other settings of the system remains the same during first scenario. Table I shows the results for increased wind uncertainty level.

For $\gamma$ between (1-2.25) there is no tripped line thus no cascading failure happens for this uncertainty range. Moving beyond $\gamma = 2.25$ the first failure in the network is observed that leads to a series of cascading failures that forms multiple islands and isolated buses. Automatic power balance on each island causes the load to be dropped to a certain level that can be supplied by generators inside the island. Successive line trips continue until all line flows drop below line thresholds because of load shedding actions. At the end of cascading failure simulation for $\gamma = 2.25$, 7941 MW (32%) of total grid load is dropped to maintain power balance of newly formed islands. Also, the first and second tripped line and their respective times are given to identify the most vulnerable lines in the network.

$\gamma$ is increased further to see the impacts of even larger uncertainty levels on severity of cascading failure. Fig. 1 shows

the total number of tripped lines and total load shedding for each wind uncertainty levels. As the $\gamma$ increases, the more lines get tripped during cascading failure which in turn leads into formation of more islands and larger load shedding as shown in Fig. 1.

TABLE I. CASCADING FAILURE RESULTS FOR FIRST SCENARIO: WIND UNCERTAINTY LEVEL

| $\gamma$ | Total trip count | # of islands formed | Total LS* (MW) | LS* (%) | First tripped line+ | Second tripped line+ |
|---|---|---|---|---|---|---|
| 2.00 | 0 | 0 | 0 | 0 | - | - |
| 2.25 | 0 | 0 | 0 | 0 | - | - |
| 2.50 | 68 | 18 | 7941 | 32 | 365 @ 54.7 | 205 @ 55 |
| 2.75 | 70 | 19 | 8344 | 33.6 | 99 @ 44.8 | 205 @ 44.9 |
| 3.00 | 71 | 19 | 8496 | 34.2 | 117 @ 44.8 | 207 @ 44.9 |
| 3.25 | 72 | 20 | 8966 | 36.1 | 365 @ 44.6 | 205 @ 45 |
| 3.50 | 74 | 21 | 9135 | 36.8 | 117 @ 30.4 | 205 @ 30.5 |
| 3.75 | 75 | 21 | 9397 | 37.8 | 115 @ 29.2 | 205 @ 29.3 |
| 4.00 | 80 | 22 | 9749 | 39.2 | 365 @ 29.2 | 205 @ 29.4 |
| 4.25 | 83 | 23 | 10005 | 40.3 | 365 @ 28.2 | 205 @ 28.3 |
| 4.50 | 83 | 23 | 10092 | 40.6 | 205 @ 28.2 | 117 @ 28.3 |
| 4.75 | 83 | 24 | 10495 | 42.3 | 365 @ 15.8 | 205 @ 16.1 |
| 5.00 | 86 | 25 | 10665 | 43 | 99 @ 15.6 | 205 @ 15.7 |

*Load Shedding
+Line number @ time [min]

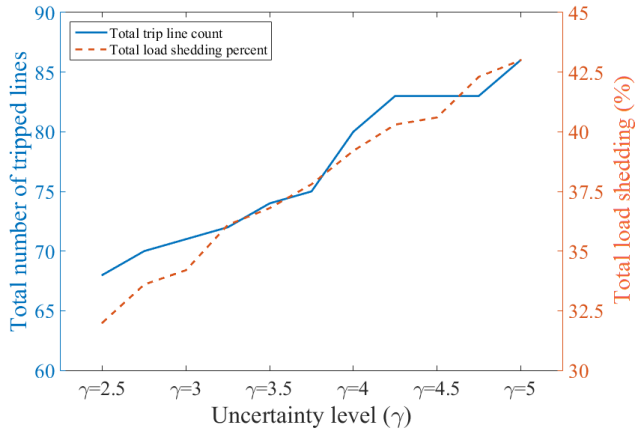

Figure 1. Total number of line trips and toal load shedding versus uncertainty level ($\gamma$).

The evolution process of cascading failures for different wind uncertainty level is shown in Fig. 2. Each evolution curve consists of two phases, the escalation phase for which the line trip rate is as high as 12 lines per minute, and the damping phase with line trip rate of approximately one line per minute. Also, it is found that as wind uncertainty level increases, the first trip happens earlier than smaller wind uncertainty cases which indicates that the minimum safety time of the entire network decreases. For example, the black bold line shows the cumulative number of line trips for uncertainty level increased by factor 5. As compared with uncertainty level increased by factor 2.5 (green line with square marker), the former results into higher number of tripped lines due to high level of wind uncertainty. Also, high uncertainty level causes contingency in multiple lines earlier compared to lower uncertainty levels. For example, the earliest cascading process is associated with the highest uncertainty level, $\gamma$=5, as indicated in Table I and happens after 15 minutes of beginning of simulation, which implies that as more uncertainty is injected to the grid, its survival time gets shorter.

The second scenario aims to see the impacts of increased penetration level of wind energy on grid vulnerability to cascading overload failures. For this scenario, wind penetration ratio is $\alpha = \frac{\sum P_{G,wind}^{max}}{\sum P_{G,total}^{max}}$ where $\sum P_{G,wind}^{max}$ is the total wind generator capacity and $\sum P_{G,total}^{max}$ is the total grid generation capacity. By replacing more conventional generators with wind generators in addition to those already installed in the network, $\alpha$ is increased to see the impacts of higher wind penetration on grid vulnerability. Note that small to medium generators are selected to be replaced with additional wind turbines to have smaller steps in $\alpha$. All other settings of the system remains the same during second scenario.

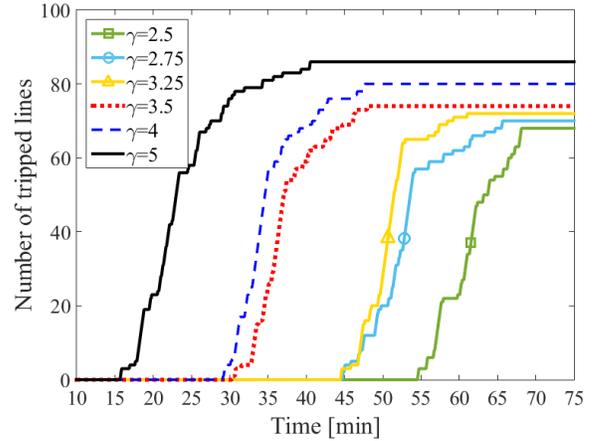

Figure 2. The evolution process of cascading failures for different wind uncertainty levels.

The results for increased wind penetration are shown in Table II. Starting from initial $\alpha = 0.036$, there is no line trip until $\alpha = 0.09$ where line 137 gets tripped at minute 2. To see further impacts of higher wind penetration, $\alpha$ is increased to 0.223 by replacing more mid-size conventional generators with wind generators. Fig. 3 shows the total number of trip line and load shedding for each penetration level $\alpha$. It is found that the higher the wind penetration ratio, the more line trip and load shedding occurs in the network.

TABLE II. CASCADING FAILURE RESULTS FOR SECOND SCENARIO: WIND PENETRATION LEVEL.

| $\alpha$ | Total trip count | # of islands formed | Total LS* (MW) | LS* (%) | First tripped line+ | Second tripped line+ |
|---|---|---|---|---|---|---|
| 0.036 | 0 | 0 | 0 | 0 | - | - |
| 0.05 | 0 | 0 | 0 | 0 | - | - |
| 0.09 | 68 | 20 | 8149 | 33 | 137 @ 2 | 101 @ 3.8 |
| 0.105 | 72 | 21 | 10156 | 41.1 | 137 @ 2 | 274 @ 3.8 |
| 0.125 | 75 | 23 | 10701 | 43.3 | 137 @ 2 | 83 @ 2.7 |
| 0.15 | 81 | 24 | 10890 | 44 | 137 @ 2 | 83 @ 2.5 |
| 0.173 | 94 | 27 | 11650 | 47.2 | 137 @ 2 | 83 @ 2.4 |
| 0.223 | 98 | 30 | 13958 | 56.5 | 137 @ 2 | 274 @ 2.3 |

*Load Shedding
+Line number @ time [min]

Fig. 4 shows the evolution process of cascading failures for different penetration levels. As shown in Table II, the triggering event for all simulation is the same (line 137), while due to random line stochastic line trip based on different uncertainty level, each one of them evolves into a different path. This potentially identifies the weakest backbone line of the network according to the new configuration of wind generators. For example, black bold line shows the cumulative number of line trips for 9% wind penetration level where starts after nearly minute 5 of simulation and stabilizes at minute 28. While, increasing wind penetration to 22.3% (green line with square marker) results into higher tripping rate, as large as 20 lines per minute during escalation phase of cascading failure, and more total number of tripped lines.

certain level, if not managed appropriately, may result into cascading failure due to considerable level of uncertainty injected and the higher the penetration the more lines get tripped and larger load shedding will be performed. Therefore, appropriate management of uncertainties via energy storage or advanced forecasting techniques is necessary for sustained growth in renewable generation. As the future extension of this study, it is of interest to incorporate AC power flow and more accurate system model to investigate voltage profiles during cascading failure evolution.

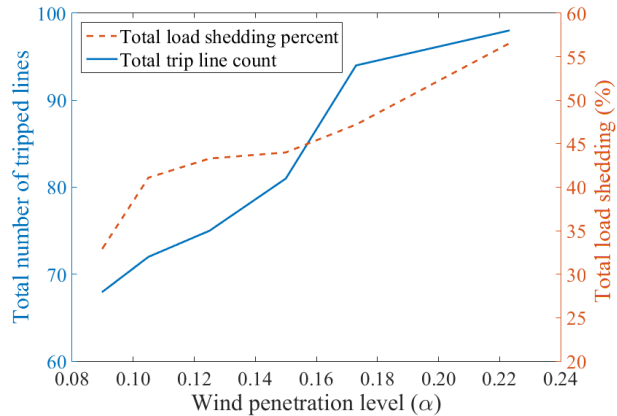

Figure 3. Total number of line trips and toal load shedding versus wind penetartion level ($\alpha$).

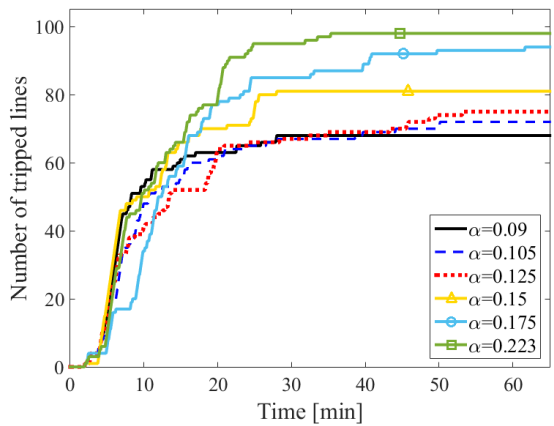

Figure 4. The evolution process of cascading failures for different wind penetartion levels.

## V. CONCLUSION

Gird vulnerability to cascading overload failures is evaluated under high penetration of wind energy. Line flow uncertainty is modeled as a zero mean signal with Gaussian distribution. Thermal stability and automatic power balance are considered in the simulation of cascading failures in IEEE 300 bus system. The DC power flow approximation is used as a simple way to calculate flow distribution of the system at each time step and find overloaded lines. Simulation results show that both high uncertainty level and high penetration level of wind generation have adverse impact on grid vulnerability to cascading failure. It is found that, higher uncertainty injected from wind due to poor forecasting results leads to severe situations in terms of total number of trip lines and load shedding. It is also found that, increasing wind penetration to a